\documentclass[final,5p,times,twocolumn]{elsarticle}
 \usepackage{graphicx}
\usepackage{bm}
\usepackage{txfonts}
\usepackage{hyperref}
\usepackage{mathrsfs}
\usepackage[english]{babel}
\usepackage{amsmath}
\usepackage{cleveref}
\usepackage[autostyle, english = american]{csquotes}
\MakeOuterQuote{"}

\date{\today}
\newcommand{\be}{\begin{eqnarray}}
	\newcommand{\ee}{\end{eqnarray}}

\newcommand{\bfk}{{\bf k}_{\perp}}

\usepackage{xcolor}

\journal{...}

\begin{document}

	
\title{Kaon structure modifications in strange hadronic matter}
	
\author{Manpreet
	Kaur}
\ead{
	ranapreeti803@gmail.com}
	
\author{Abi Jebarson A
}
\ead{abijebarson@gmail.com}
	
\author{Dhananjay Singh}
\ead{
	snaks16aug@gmail.com}
	
\author{Navpreet Kaur}
\ead{knavpreet.hep@gmail.com}

\author{Suneel Dutt}
\ead{dutts@nitj.ac.in}
\author{Harleen Dahiya}
\ead{dahiyah@nitj.ac.in}
\author{Arvind Kumar}
\ead{kumara@nitj.ac.in}

\affiliation{Computational High Energy Physics Lab, Department of Physics, Dr. B.R. Ambedkar National
		Institute of Technology, Jalandhar, Punjab, 144008, India}
	
%
\begin{abstract}
We present the valence quark distributions of the kaons in an isospin asymmetric dense strange medium consisting of nucleons and hyperons. The comparative analysis of in-medium parton distribution functions, electromagnetic form factors, and charge densities with respect to the free space distributions is studied in the light-cone quark model. The medium effects are incorporated in these distribution functions by using the effective quark masses, computed from the chiral SU(3) quark mean field model for finite values of baryonic density, isospin asymmetry, and strangeness fraction parameters. We observe a suppression of the kaon electromagnetic form factors and a redistribution of charge density in high-density strange matter.
\end{abstract}
%
\maketitle
%
%
\section{Introduction\label{secintro}}
The comprehensive information about the internal structure of the hadrons can be inferred from their valence quark distribution functions in momentum and impact parameter space. At the partonic level, one-dimensional distribution functions, such as parton distribution functions (PDFs) and electromagnetic form factors (EMFFs), serve as a vital source of information. The former describes the probability of locating a valence quark with longitudinal momentum fraction $x$ \cite{Collins:1981uw, Martin:2009iq}, whereas the latter signifies the charge distribution of a valence quark forming the particular hadron. \cite{Perdrisat:2006hj, Diehl:2013xca, Khodjamirian:2006st}. Fourier transformations of the EMFFs of hadrons lead to the extraction of charge and magnetization densities in the impact parameter space \cite{Miller:2010tz, Miller:2009sg, Kim:2008ghb}. The valence quark distributions are expected to be modified when a hadron is immersed in the nuclear medium. Experimentally, such modification in the structure functions of the bound hadrons in nuclei compared to the free space has been observed, and is referred to as the EMC effect \cite{EuropeanMuon:1983wih}, suggesting the influence of the nuclear medium on the quark dynamics. The interpretation of the EMC effect measurements for valence quarks in terms of nucleon--nucleon short-range correlations (SRC) in nuclei is also explored \cite{Hen:2016kwk}.

Taking the observation of the EMC effect into account, in recent years, distribution amplitudes (DAs), PDFs, EMFFs and transverse momentum-dependent distributions have been explored in symmetric and isospin asymmetric nuclear medium for the pseudoscalar mesons \cite{Hutauruk:2021kej, Arifi:2023jfe, Arifi:2024tix, Hutauruk:2018qku, Hutauruk:2019was, Hutauruk:2019ipp,Yabusaki:2023zin, Er:2022cxx, Puhan:2024xdq, Kaur:2024wze, Singh:2024lra}. In Ref. \cite{Arifi:2024tix}, in-medium EMFFs of light and heavy-light pseudoscalar mesons were studied, and it was found that EMFFs of pions and kaons undergo more significant modifications than heavy-light mesons. Pion \cite{Hutauruk:2018qku} and kaon \cite{Hutauruk:2019was} EMFFs have been explored in the hybrid framework of the Nambu–Jona–Lasinio (NJL) and quark--meson coupling (QMC) model. The in-medium EMFFs for both pseudoscalar mesons are found to be suppressed compared with those in free space as the nuclear medium density increases. In Ref. \cite{Hutauruk:2019ipp}, the same hybrid approach was employed to study the PDFs and EMFFs of the pion and kaon in a symmetric nuclear medium. In addition to this, kaon properties have been studied in the symmetric nuclear matter using the light-front constituent quark model combined with the QMC model and the Bethe-Salpeter approach (BSA) in Ref. \cite{Yabusaki:2023zin}, and in the context of QCD sum rules in Ref. \cite{Er:2022cxx}. In an isospin asymmetric nuclear medium, the impact of the isospin asymmetry parameter and finite temperature has been explored for the case of pions and kaons \cite{ Puhan:2024xdq, Kaur:2024wze, Singh:2024lra}. In this work, we present the dependence of PDFs, EMFFs, and charge densities on the strangeness fraction parameter $f_s$ for kaon $K^+$ immersed in a medium having both nucleons and hyperons, relevant for heavy-ion collisions (HICs). The medium produced in HICs consists not only of nuclear matter but also strange matter, which modifies the relevant physical observables.

The enhancement in the strange-to-non-strange hadron ratio with the increase in charged-particle multiplicity density serves as a key indicator of the strange medium generated in $pp$ and $ p$–Pb collisions, as observed in ALICE data \cite{vanWeelden:2024xyy}.
In HICs, the initial system maintains strangeness neutrality, which requires that $s$ and $\bar s$ quarks are created in pairs. These quarks subsequently form strange (anti-strange) mesons and baryons or are integrated into mesons with hidden strangeness \cite{Song:2022jcj}.
Strange hadrons are particularly sensitive to the medium produced in HICs \cite{KAOS:2000ekm, STAR:2024znc}.
This sensitivity is further highlighted by the observation that among the three light quark flavors, the $s$ quark flavor undergoes greater mass modifications than the $u$ and $d$ quarks as a function of $f_s$ \cite{Dutt:2024lui}. Therefore, we consider the lightest strange pseudoscalar meson, i.e., kaon $K^+$, to investigate the influence of $f_s$ on the PDFs, EMFFs, and charge densities within a strange matter at zero temperature. The concept of zero temperature is particularly significant in the analysis of massive neutron star cores, where strange hadrons contribute to a system capable of sustaining phase coexistence under near-zero temperature conditions \cite{Glendenning:1992vb}. The future Electron-Ion Collider (EIC) aims to explore in-medium mesons under zero-temperature conditions.

The distribution functions are investigated in the light-cone quark model (LCQM), and the inputs required to study the medium effects are computed from the chiral SU(3) quark mean field (CQMF) model. The light-cone formalism offers an effective approach to study the relativistic effects of quark dynamics in hadrons \cite{Dirac:1949cp}. 
LCQM has successfully explained the multidimensional structure of the kaon in free space through the study of generalized transverse momentum-dependent parton distributions, generalized parton distributions (GPDs), and possible spin-orbit correlations in the context of Wigner distributions \cite{Kaur:2019jow}. For a kaon within a strange medium, the effective quark masses are calculated within the CQMF model, which treats quarks as degrees of freedom and binds them within a hadron via a confining potential. This model has been used to study the properties of both nuclear and strange hadronic matter \cite{Wang:2001jw, Wang:2001hw}. Additionally, the magnetic moments of baryons at finite temperatures have been extensively explored both in nuclear and strange medium \cite{Singh:2016hiw, Singh:2018kwq, Kumar:2023owb}.
The hybrid approach of the LCQM with the CQMF model has been applied to study the medium-modifications of kaons within both isospin symmetric and asymmetric nuclear medium \cite{Singh:2024lra, Pandey:2025rqo}.
However, to our knowledge, the valence quark properties of kaons in a strange medium have not yet been explored. In this work, we aim to explore these in-medium properties of kaon in an isospin asymmetric strange medium using the same approach (LCQM + CQMF).

The paper is organized as follows. In Sec. \ref{Method}, the detailed description of both models is presented.
  Sec. \ref{results} provides an analysis of both free space and in-medium PDFs. Additionally, this section presents findings of EMFFs and charge density. Finally, we summarize our results in Sec. \ref{summary}.

\section{Methodology}
\label{Method}

To investigate the medium modifications of kaons in strange hadronic medium, we employ the LCQM, a relativistic framework that enables a nonperturbative treatment of valence quark dynamics and has been widely successful in describing mesonic observables such as PDFs, DAs, and GPDs \cite{Kaur:2020vkq}. Medium effects arising due to finite baryon density, isospin asymmetry, and strangeness fraction are incorporated by using effective quark masses computed from the CQMF model. These in-medium quark masses serve as fundamental inputs to the LCQM for calculating PDFs, EMFFs, and charge densities.

In the light-front frame, the kaon is represented as a superposition of Fock states, dominated by the leading two-particle state consisting of a quark and an antiquark. The state with longitudinal spin projection $S_z=0$ can be written as \cite{Lepage:1980fj,Qian:2008px}
\begin{align}
|K(P^+, \mathbf{P}_\perp)\rangle = \sum_{\lambda_1, \lambda_2} \int \frac{dx d^2\mathbf{k}_\perp}{16\pi^3\sqrt{x(1-x)}} \, \psi_K^{\lambda_1\lambda_2}(x, \mathbf{k}_\perp) \nonumber \\
|x,\mathbf{k_\perp,\lambda_1,\lambda_2})\rangle,
\end{align}
where $x$ is the longitudinal momentum fraction carried by the quark and $\mathbf{k}_\perp$ is its transverse momentum. $\lambda_1$ and $\lambda_2$ are the helicities of the quark and the antiquark in the meson, respectively. The total light-cone wave function is decomposed into spin and momentum space components as 
\begin{equation}
\psi_K^{\lambda_1\lambda_2}(x, \mathbf{k}_\perp) = \Phi^{\lambda_1 \lambda_2}_K(x, \mathbf{k}_\perp) \, \varphi_K(x, \mathbf{k}_\perp),
\end{equation}
with $\Phi_K^{\lambda_1\lambda_2}(x, \mathbf{k}_\perp)$ as the spin and $\varphi_K(x, \mathbf{k}_\perp)$ being the momentum space wave function. The meson light-cone spin wavefunctions are \cite{Qian:2008px} 

\begin{eqnarray}
\Phi^{\uparrow,\uparrow}_K(x,{\bf k}_\perp)&=&-\frac{1}{\sqrt{2}}\frac{k^1-i k_{}^2}{\sqrt{{\bf k}^2_\perp+l^2}},\nonumber\\
\Phi^{\uparrow,\downarrow}_K(x,{\bf k}_\perp)&=&\frac{1}{\sqrt{2}}\frac{(1-x)m^*_q+x m^*_{\bar{q}}}{\sqrt{{\bf k}^2_\perp+l^2}},\nonumber\\
\Phi^{\downarrow,\uparrow}_K(x,{\bf k}_\perp)&=&-\frac{1}{\sqrt{2}}\frac{(1-x)m^*_q+x m^*_{\bar{q}}}{\sqrt{{\bf k}^2_\perp+l^2}},\nonumber\\
\Phi^{\downarrow,\downarrow}_K(x,{\bf k}_\perp)&=&-\frac{1}{\sqrt{2}}\frac{k^1+i k_{}^2}{\sqrt{{\bf k}^2_\perp+l^2}}, \label{wavefunctions}
\end{eqnarray}
with
\begin{eqnarray}
 l^2=(1-x)m_q^{*2}+x m_{\bar{q}}^{*2}-x(1-x)(m^*_q-m^*_{\bar{q}})^2.
\end{eqnarray}
Here, for the kaon, $k^1$ and $k_{}^2$ represent the transverse components of momenta for the constituent quark or antiquark. The four-vector momenta for the constituent valence quark and antiquark are given by
\begin{align}
k_q &= \left(x P^+, \frac{\mathbf{k}_\perp^2 + m_q^{*2}}{x P^+}, \mathbf{k}_\perp \right), \nonumber \\
k_{\bar{q}} &= \left((1-x) P^+, \frac{\mathbf{k}_\perp^2 + m_{\bar{q}}^{*2}}{(1-x) P^+}, -\mathbf{k}_\perp \right),
\end{align}
respectively. The momentum-space wave function of the kaon is modeled using the Brodsky–Huang–Lepage (BHL) prescription and takes the form \cite{Xiao:2002iv, Yu:2007hp}:
\begin{align}
\varphi_K(x, \mathbf{k}_\perp) = \mathcal{A} \exp\Biggl[ 
- \frac{1}{8\beta_K^2} \left( 
\frac{m_q^{*2} + \mathbf{k}_\perp^2}{x} 
+ \frac{m_{\bar{q}}^{*2} + \mathbf{k}_\perp^2}{1 - x} \right) \nonumber \\
- \frac{\left(m_q^{*2} - m_{\bar{q}}^{*2}\right)^2}
{8\beta_K^2 \left( 
\frac{m_q^{*2} + \mathbf{k}_\perp^2}{x} 
+ \frac{m_{\bar{q}}^{*2} + \mathbf{k}_\perp^2}{1 - x} 
\right)} 
\Biggr],
\label{eq:wave function_BHL}
\end{align}
where $\mathcal{A}$ is a normalization constant, and $\beta_K$ is the harmonic scale parameter determined by fitting to the kaon decay constant. Here, we used $\beta_K = 0.393$ and $\mathcal{A}$ is calculated for each set of masses used. The effective mass of a constituent quark in the medium $m_q^*$ is modified by scalar meson fields and, in the CQMF model, is expressed as \cite{Wang:2002pza}
\begin{equation}
    m^*_q = -g_{q\sigma}\sigma - g_{q\zeta}\zeta - g_{q\delta}I^q_3\delta + m_q^0,
    \label{mass}
\end{equation}
where $g_{qS}$ are quark coupling constants for the scalar fields $S=\sigma,\zeta,\delta$. $I_3^q$ is the third component of isospin for quark, with values $I_3^u=+\frac{1}{2}, I_3^d=-\frac{1}{2}$ and $I_3^s = 0$. The additional mass term used to reproduce the empirical vacuum masses of quarks is taken as $m_{u,d}^0=0$ and $m_s^0 = 77$ MeV. The effective quark energy in the medium includes contribution from the vector meson fields and is given by
\begin{equation}
    e_q^* = e_q - g_{q\omega}\omega - g_{q\rho}I_3^q\rho - g_{q\phi}\phi.
    \label{energy}
\end{equation}
Here, $g_{qV}$ are quark couplings for the vector fields $V=\omega,\rho,\phi$. For a strange, isospin-asymmetric hadronic medium at temperature $T=0$ and finite baryon density, the thermodynamic potential in the grand canonical ensemble is defined as \cite{Wang:2005vg}
\begin{align}
\Omega &= - \sum_j \frac{\gamma_j}{48\pi^2} \left[
\left( 2k_{Fj}^3 - 3{M_j^*}^2 k_{Fj} \right) v_j^* \right. \notag \\
&\quad \left. + 3{M_j^*}^4 \ln\left( \frac{k_{Fj} + v_j^*}{M_j^*} \right)
\right] - \mathcal{L}_M - \mathcal{V}_{\text{vac}},
\end{align}
where the summation runs over baryon species $(p,n,\Lambda,\Sigma^{\pm,0}, \Xi^{-,0})$ and $\gamma_j$ is the degeneracy factor of baryons, which is equal to 2 for octet baryons. The term ${\cal L}_{M}$ includes scalar, vector meson self-interactions and explicit symmetry breaking of the interaction Lagrangian of CQMF model. The term ${\cal{V}}_{\text{vac}}$ is subtracted to acquire zero vacuum energy. $M^*_j$, $k_{Fj}$, and $\nu_j^*$ are the effective baryon mass, fermi momentum, and chemical potential of baryons, respectively. The scalar and the vector fields are determined by solving the set of coupled nonlinear equations obtained by minimizing $\Omega$ with respect to each field. These fields depend on the baryon density $\rho_B$, isospin asymmetry $\eta =-\frac{\sum_j I_{3}^{j} \rho^{v}_{j}}{\rho_{B}}$, and strangeness fraction $ f_s=\frac{\sum_j \vert S_{j} \vert \rho^{v}_{j}}{\rho_{B}}$, with $\rho_j^v$ and $S_j$ being the vector density and strangeness quantum number of the baryon $j$. 

Using the in-medium masses from Eq. (\ref{mass}), the unpolarized valence quark PDF $f^q_K(x,m_q^*,m_{\bar{q}}^*)$ and hereafter referred to simply as $f^q_K(x)$, is computed as \cite{Maji:2016yqo}
\begin{equation}
\label{pdf_1}
f^q_K(x) = \int \frac{d^2 \mathbf{k}_\perp}{16\pi^3} \sum_{\lambda_1, \lambda_2} 
|\psi_K^{\lambda_1, \lambda_2}(x, \mathbf{k}_\perp)|^2,
\end{equation}
which reflects the probability of finding a quark with momentum fraction $x$ inside the kaon. 
Using the explicit spin structure of the kaon wave function in Eq. (\ref{wavefunctions}), the valence quark PDF takes the form

\begin{align}
    f^{q}_K(x)&=&\int \frac{d^2 {\bf k}_\perp}{16\pi^3}\bigg[{\bf k}_\perp^2+{((1-x)m_q^* + x m_{\bar{q}}^*)^{2}}\bigg]
 \frac{|\varphi(x,{\bf k_\perp})|^2}{{\bf k}_\perp^2+{l}^2}.\label{pdf_2}
\end{align}

The EMFF of the quark/antiquark, $F_K^{q/\bar{q}}(Q^2)$, encodes the distribution of electric charge and is evaluated using the GPDs at zero skewness ($\xi = 0$). The quark-quark correlator to evaluate GPD is as follows
\cite{Kaur:2018ewq}

\begin{eqnarray}
    H_K^u(x, \xi=0, q^2)=\frac{1}{2} \int \frac{dz^-}{2\pi} e^{ik\cdot z} \langle K^+(P_f)|\bar{\psi} (-z/2) \nonumber \\
    \gamma^+ \psi(z/2) |K^+ (P_i) \rangle |_{z^+=0, \bf{z}_\perp=0},
\end{eqnarray}
where $(-q^2)=Q^2$ is in the units of GeV$^2$. The quantities $P_f$ and $P_i$ are the final and initial state momentum of a kaon. The zeroth moment of these unpolarized GPDs $H_K^u(x, \xi=0, q^2)$ gives the EMFF \cite{Kaur:2018ewq}, and the expression for EMFF in terms of the overlap form of LCWFs can be written as
 \begin{align}
 \label{FF}
     F^q_K(Q^2) = \int \frac{dx~d^2 \mathbf{k}_\perp}{16\pi^3} \sum_{\lambda_1, \lambda_2} \psi_K^{*~\lambda_1, \lambda_2}(x, \mathbf{k}_\perp'')~\psi_K^{\lambda_1, \lambda_2}(x, \mathbf{k}_\perp'),
 \end{align}
where final and initial state momentum of constituent valence quark of a kaon is expressed by $\bfk^{\prime \prime}=\bfk-(1-x)q/2$ and $\bfk^{\prime}=\bfk+(1-x)q/2$, respectively. 
Upon substituting the decomposed spin wavefunctions from Eq. (\ref{wavefunctions}) and momentum wavefunctions into the above expression for EMFF, it takes the form

\begin{eqnarray}
 \label{FF2}
    F^{q}_K(Q^2)&=&\int \frac{dx~d^2 {\bf k}_\perp}{16\pi^3}\bigg[{\bf k}_\perp^2 - (1-x)^2\frac{Q^2}{4} \\ \nonumber &+&{((1-x)m_q^* +  x m_{\bar{q}}^*)^2}\bigg]
 \frac{\varphi^*(x,{\bf k''_\perp})\varphi(x,{\bf k'_\perp})}{\sqrt{{\bf k''}_\perp^2+{l}^2}\sqrt{{\bf k'}_\perp^2+{l}^2}}.
\end{eqnarray}

The corresponding antiquark PDFs and EMFF can be obtained by swapping the quark and antiquark masses, i.e., $m_q^* \leftrightarrow m^*_{\bar{q}}$ in Eq. (\ref{pdf_2}) and Eq. (\ref{FF2}), respectively. This gives $f^{\bar{q}}_K(x)=f^q_K(x, m^*_{\bar{q}}, m^*_q)$ and $F^{\bar{q}}_K(Q^2)=F^q_K(Q^2, m^*_{\bar{q}}, m^*_q)$. In the light-cone formalism, the GPDs reduce to overlap integrals of LCWFs, thereby incorporating medium effects via the effective quark masses \cite{Kaur:2024wze}. 

The transverse spatial charge distribution of quark/antiquark is obtained via a two-dimensional Fourier transform of the EMFF as 
\cite{Miller:2009sg}
\begin{equation}
\label{charge_density}
\rho^{q/\bar{q}}_K(b_\perp) = \frac{1}{2\pi} \int_0^\infty dQ\, Q\, J_0(b_\perp Q)\, F_K^{q/\bar{q}}(Q^2),
\end{equation}
where $b_\perp$ is the transverse impact parameter and $J_0$ is the zeroth-order Bessel function of the first kind. The charge density represents the transverse charge density of a valence quark at distance $b_\perp$ from the center of momentum of the kaon. The in-medium effective masses $m_q^*$ and $m_{\bar{q}}^*$ influence $F_K^{q/\bar{q}}(Q^2)$ via the wave functions, and thus modify the spatial profile of $\rho^{q/\bar{q}}_K(b_\perp)$. It provides a detailed picture of the medium-induced redistribution of charge in the kaon's internal structure.

\begin{table}
\centering
\def\arraystretch{1.8}
\begin{tabular}{ccccccccc}
\hline \hline
 Quarks &${g_{\sigma i}}$& ${g_{\zeta i}}$  &   ${g_{\delta i}}$ &   ${g_{\omega i}}$ &${g_{\rho i}}$ &${g_{\phi i}}$\\ \hline
$u$&2.72\ & 0 &  2.72 &  3.23&  3.23 &  0\\ 
$d$&2.72 & 0 & 2.72 & 3.23  & 3.23 & 0 \\ 
$s$&0& 3.847 & 0 & 8.89 & 0& 4.57\\ 

\hline \hline
\end{tabular}
\caption{Coupling constants of scalar ($\sigma, \zeta$ and $\delta$) and vector ($\omega, \rho$ and $\phi$) fields with quark flavors.}
\label{coupling}
\end{table}

\section{Results and discussion}
\label{results}
\begin{figure*}
\centering
\includegraphics[width=18cm]{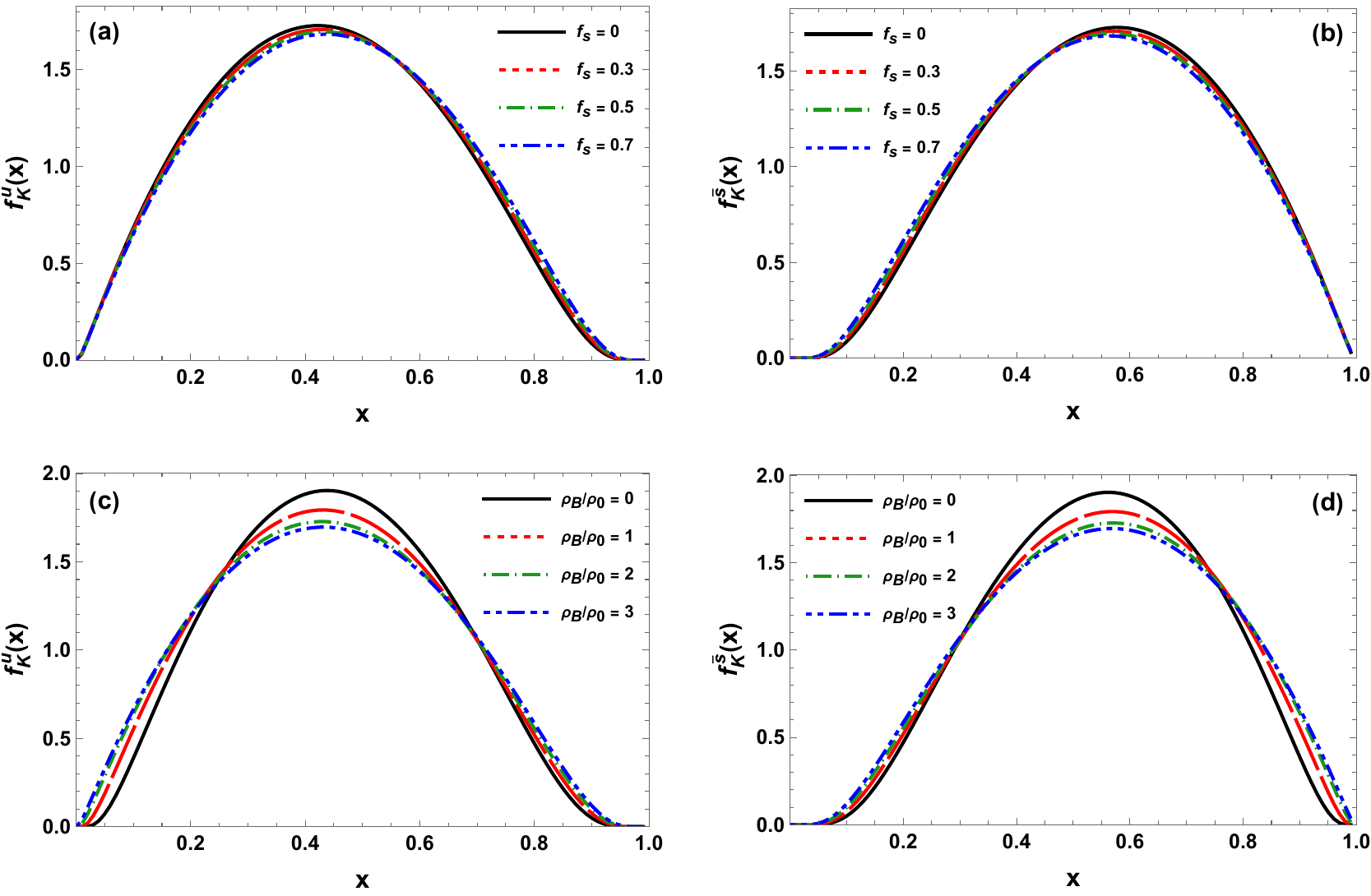}
\caption{PDFs of the $u$ quark in the left panel and $\bar{s}$ antiquark in the right panel are plotted as a function of longitudinal momentum fraction $x$. The results are shown for different values of strangeness fraction $f_s = 0, 0.3,0.5,0.7$, at fixed baryon density ratio $\rho_B/\rho_0 = 3$ and isospin asymmetry $\eta = 0.5$ in subplots (a) and (b), whereas for a range of $\rho_B/\rho_0 = 0, 1, 2, 3$ and keeping $f_s = \eta = 0.5$  fixed in subplots (c) and (d).}   
\label{fig_PDF}
\end{figure*}

In this section, we discuss the results of PDFs and EMFFs of valence quark and antiquark of the kaon in LCQM, using in-medium quark masses as determined by the CQMF model. Specifically, we investigate how the presence of hyperons $\Sigma$, $\Lambda$, and $\Xi$, along with nucleons, within the medium influences these properties of the kaons. The CQMF model calculates effective quark masses using medium-modified scalar fields ($\sigma$, $\zeta$, $\delta$), as described in Eq. (\ref{mass}). These medium-modified quark masses $m^*_{i}$ serve as inputs in Eqs. (\ref{pdf_1}),     (\ref{FF}) and (\ref{charge_density}) within the LCQM.
The coupling constants that define the interactions between mesonic fields and quark flavors are listed in Table \ref{coupling}. The detail of parameters of the CQMF model can be found in Refs. \cite {Singh:2024lra,Wang:2002pza}.
In Fig. \ref{fig_PDF}, the valence PDFs of the $u$ quark (left panel) and $\bar s$ antiquark (right panel) in the kaon within asymmetric strange matter are plotted as a function of longitudinal momentum fraction $x$.
Figures \ref{fig_PDF}(a) and (b) show the results
 over a finite range of strangeness fraction $f_s$, with
 fixed baryonic density ratio $\rho_B/\rho_0 = 3$ ($\rho_0=0.16$ fm$^{-3}$) and isospin asymmetry $\eta = 0.5$. 
  In Fig. \ref{fig_PDF}(a), the value of $f^u_{K}(x)$ within the medium decreases as the strangeness fraction $f_s$ increases from 0 to 0.7 at lower $x$ ($x< 0.6$). Conversely, at higher values of $x$ ($x>$0.6), the PDF value increases with increasing $f_s$. For antiquark ($\bar s$), as expected, we observe the opposite trend in the PDF $f^{\bar s}_{K}(x)$, as shown in Fig. \ref{fig_PDF}(b). 
Moreover, the position of the PDF peak is observed to shift toward higher longitudinal momentum fraction $x$ in Fig. \ref{fig_PDF}(a) as the medium strangeness increases, and a similar shift is also noticed in $f^{\bar s}_{K}(x)$, but towards lower $x$ as depicted in Fig. \ref{fig_PDF}(b). In subplots (c) and (d) of Fig. \ref{fig_PDF}, the calculated PDFs are shown over a range of baryonic density ratios ($\rho_B/\rho_0=0,1,2,3$) as a function of $x$ for $u$ and $\bar s$ quarks, respectively, at fixed isospin asymmetry parameter  and strangeness fraction ($\eta = f_s = 0.5$). With increasing $\rho_B/\rho_0$, the PDFs flatten and their longitudinal momentum fraction dependence broadens, as seen in Figs. \ref{fig_PDF}(c) and \ref{fig_PDF}(d). 
The distribution $f^{u}_{K}(x)$ broadens more at lower $x$, and an opposite behavior of $f^{\bar s}_{K}(x)$ is observed as it becomes broader at higher values of $x$. This in-medium modification of the PDFs relative to their vacuum values is attributed to a more pronounced reduction in the quark mass $ m^*_{i}$ in strange matter compared to nuclear matter, reflecting the influence of partial restoration of chiral symmetry. The concept of partial restoration of chiral symmetry has also been supported by the observations from the low--energy pion--nucleus scattering \cite{Friedman:2004jh}, the deeply bound pionic atoms \cite{Suzuki:2002ae}, as well as di--pion production in hadron--nucleus and photon-nucleus reactions \cite{CHAOS:1996nql, CHAOS:2004rhl}. 
\par
\begin{figure*}
\centering
\includegraphics[width=18cm]{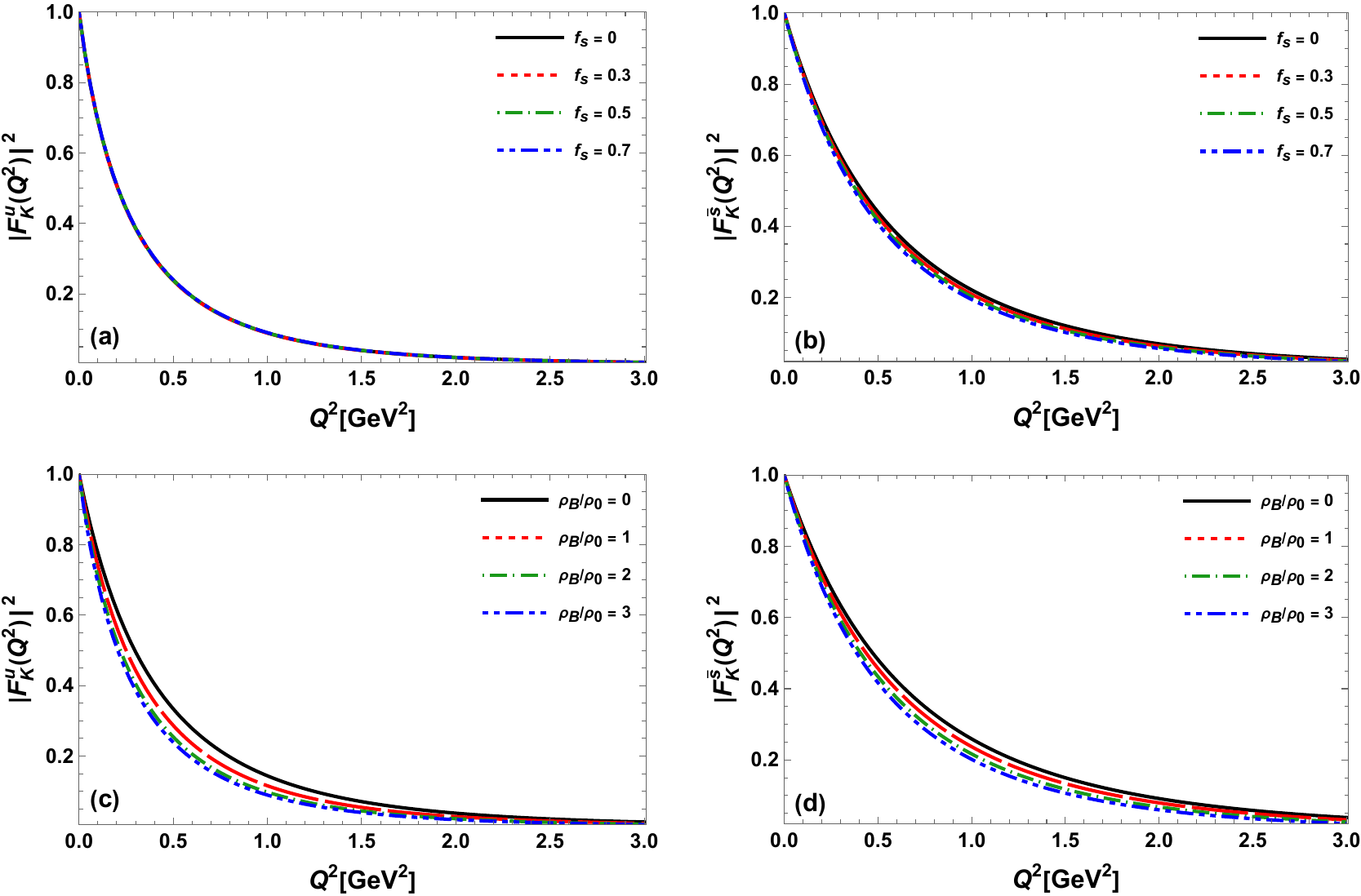}
\caption{EMFF of the $u$ quark and the $\bar{s}$ antiquark are plotted as a function of momentum transfer $Q^2$ (GeV$^2$). The subplots (a) and (b) show the results at strangeness fraction values $f_s = 0, 0.3,0.5,0.7$ and fixed baryonic density ratio $\rho_B/\rho_0 = 3$. The subplots (c) and (d) compare the EMFF when $\rho_B/\rho_0 = 0, 1, 2, 3$, for fixed $f_s = \eta = 0.5$.}
\label{fig_EFF}
\end{figure*}
\begin{figure*}
\centering
\includegraphics[width=18cm]{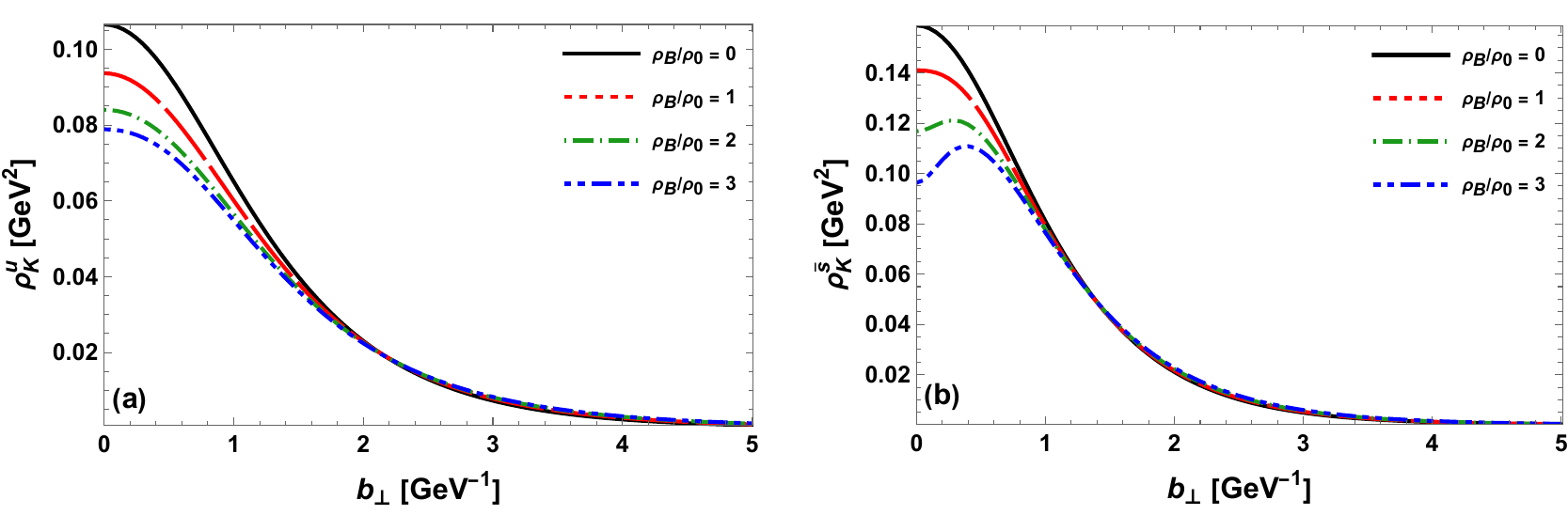}
\caption{The charge density of the $u$ quark $\rho_K^u$ in the left panel and $\bar{s}$ antiquark $\rho_K^{\bar{s}}$ in the right panel are plotted as a function of the impact parameter $b_{\perp}$ (GeV$^-1$), for baryon density ratios $\rho_B/\rho_0 = 0, 1, 2, 3$, for fixed $f_s = \eta = 0.5$.}
\label{fig_CDs1}
\end{figure*}
\begin{figure*}
\centering
\includegraphics[width=18cm]{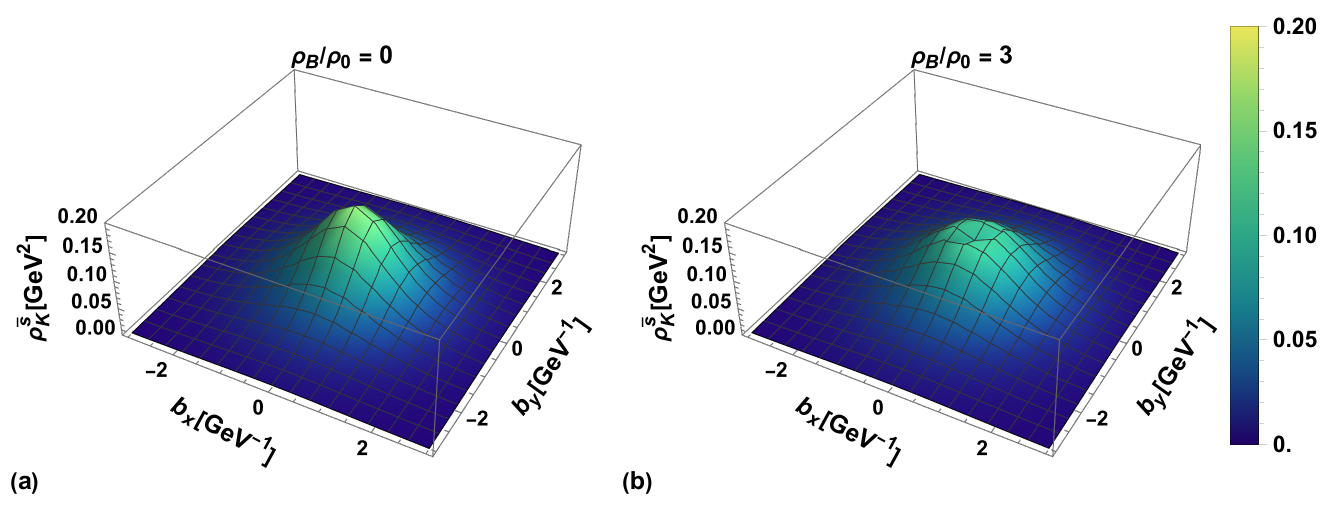}
\caption{The charge density distribution of the $\bar{s}$ antiquark is plotted as a function of the two-dimensional impact parameter space ($b_x,b_y$) at $\rho_B/\rho_0 = 0$ in the left panel and $\rho_B/\rho_0 = 3$ in the right panel, for fixed $f_s = \eta = 0.5$.}
\label{fig_CDs}
\end{figure*}

The EMFFs of the $u$ quark, $|F^{{u}}_{K}(Q^2)|^2$, and the $\bar{s}$ antiquark, $|F^{{\bar s}}_{K}(Q^2)|^2$, are numerically calculated from the GPD expression as defined in Eq. (\ref{FF}). In Fig. \ref{fig_EFF}, we present the results for the EMFFs of the $u$ and $\bar{s}$ quarks in the kaon within strange matter as a function of momentum transfer $Q^2$, for various values of strangeness fraction, $f_s$ and baryon density ratio, $\rho_B/\rho_0$. In Figs. \ref{fig_EFF}(a) and \ref{fig_EFF}(b), the impact of medium's strangeness on these EMFFs is examined at a fixed $\rho_B/\rho_0 = 3$ and $\eta = 0.5$.
 For lower values of $\rho_B/\rho_0$, the EMFFs of both quarks exhibit minimal variation with respect to $f_s$. Therefore, these results are not shown in the present work. Even at $\rho_B/\rho_0 = 3$, the EMFF $|F^{{u}}_{K}(Q^2)|^2$ shows negligible change as $f_s$ increases, as seen in Fig. \ref{fig_EFF}(a). However, $|F^{{\bar s}}_{K}(Q^2)|^2$ exhibits a more significant reduction with increasing $f_s$, as expected, and is illustrated in Fig. \ref{fig_EFF}(b).
Subplots (c) and (d) of Fig. \ref{fig_EFF} show the behavior of $u$ and $\bar{s}$ quark form factors, respectively, as a function of $Q^2$ for different values of $\rho_B/\rho_0$, at fixed $f_s = \eta = 0.5$. A more noticeable drop in the form factors is observed as $\rho_B/\rho_0$ increases.
According to Ref. \cite{Yabusaki:2023zin}, the EMFF $|F^{{\bar s}}_{K}(Q^2)|^2$ remains unchanged in the presence of nuclear matter as $\rho_B/\rho_0$ increases. However, our findings show a significant variation in strange matter, as illustrated in Fig. \ref{fig_EFF}(d). This variation can be attributed to the direct interaction between the strange quark in the kaon and the surrounding strange medium. 
Furthermore, when the Fourier transform of the EMFFs is evaluated to obtain the charge density using Eq. (\ref{charge_density}), a notable distinction between the quark and antiquark is observed, as illustrated in Fig. \ref{fig_CDs1}.
The $u$ quark charge density $\rho^u_{K}$ shows the charge being concentrated at the core, as expected for an elementary charged particle and most hadronic particles, as shown for the pion in Ref. \cite{Miller:2010tz}. The $\bar{s}$ antiquark exhibits a similar distribution at $\rho_B/\rho_0=0 $. At higher baryonic density ratios, however, the charge of the $\bar s$ antiquark is no longer centralized resulting in a region of reduced charge at the core. To enhance the visualization of $\bar s$ antiquark charge density $\rho^{\bar s}_{K}$, we illustrate the 3D surface plot in Fig. \ref{fig_CDs}. In this figure, the charge density of $\bar s$ antiquark in the kaon is plotted as a function of impact parameter coordinates $b_x$ and $b_y$, for $\rho_B/\rho_0=0 $ in the left panel and $\rho_B/\rho_0=3 $ in the right panel, at fixed $f_s = 0.5$ and $\eta = 0.5$.
In Fig. \ref{fig_CDs}(a), a higher density core is observed at lower baryonic density, whereas when the baryonic density increases, the charge density at the core decreases, as shown in Fig. \ref{fig_CDs}(b). 
The in-medium modification of the antikaon $K^-$ properties in strange matter exhibits a trend similar to that observed for the $K^+$. Therefore, this study focuses on presenting results related to $K^+$ only.

\section{Summary}
\label{summary}
To conclude, in this study, we have explored the medium modification of kaon properties in isospin asymmetric strange matter using a combined approach of LFQM and CQMF model. We conducted a comparative analysis of medium-modified PDFs, EMFFs, and charge densities of $u$ and $\bar s$ quarks inside the kaon, relative to their free-space distributions. Within the CQMF framework, the in-medium quark masses were computed through the exchange of scalar fields $\sigma$, $\zeta$, and $\delta$. These effective quark masses serve as inputs for the calculations of PDFs and EMFFs of the quark and antiquark in the kaon. The observed alterations in the properties of the kaon can be attributed to the reduction in quark masses, suggesting a partial restoration of chiral symmetry.
We observed that the in-medium PDFs of the $u$ quark decrease while those of the $\bar s$ antiquark increase, as the strangeness fraction of the medium increases, specifically for the lower range of longitudinal momentum fraction at fixed baryonic density. As the $\rho_B/\rho_0$ ratio rises, the PDFs become flatter, and their dependence on the longitudinal momentum fraction becomes wider. Furthermore, the medium-modified EMFFs of the kaon's constituents exhibit a more pronounced decline with increasing $\rho_B/\rho_0$ compared to increasing $f_s$, i.e., the EMFFs display a stronger correlation with change in baryonic density than the strangeness of the medium.
\par In addition, we presented results for the charge densities of $u$ and $\bar s$ quarks in the kaon. We observed that at higher baryonic densities, the charge distribution of the $\bar s$ antiquark is no longer centralized at zero impact parameter. Instead, as the impact parameter space increases, a notable reduction in its value is observed, similar to the case of lower baryonic density.
Our findings suggest that the observed changes in the PDFs and EMFFs, as well as the charge densities of the $u$ and $\bar s$ quarks within the kaon are likely due to modifications in its internal structure and interactions with the surrounding medium.
The properties of kaons, when interacting with a medium, are expected to yield significant insights for future kaon findings at the EIC \cite{AbdulKhalek:2021gbh} and J-PARC \cite{Lim:2011zza}.

\section*{\bf Acknowledgement}
H.D. would like to thank  the Science and Engineering Research Board, Anusandhan-National Research Foundation, Government of India under the scheme SERB-POWER Fellowship (Ref No. SPF/2023/000116) for financial support. A. K. sincerely acknowledge Anusandhan National
Research Foundation (ANRF), Government of India for
funding of the research project under the Science and
Engineering Research Board-Core Research Grant
(SERB-CRG) scheme (File No. CRG/2023/000557).

\end{document}